\documentclass[10pt, conference, letterpaper]{IEEEtran}

\usepackage{tikz}

\usepackage{filecontents,cite}
\usepackage{booktabs} 
\usepackage{graphicx, epsfig, color, amsmath}
\usepackage[ruled,vlined]{algorithm2e}
\usepackage[]{subfigure}
\usepackage{float}
\usepackage{comment}
\usepackage{authblk}

\author[1]{Joshua Morris}
\author[1]{Dan Lin}
\author[2]{Marcellus Smith}

\affil[1]{University of Missouri, Columbia, MO}
\affil[2]{Auburn University, Auburn, AL}

\begin{document}

  \title{\large Fight Virus Like a Virus: A New Defense Method Against File-Encrypting Ransomware  }
  \maketitle
  \author{\vspace{30pt}}

  \begin{abstract}


Nowadays ransomware has become a new profitable form of attack. This type of malware acts as a form of extortion which encrypts the files in a victim's computer and forces the victim to pay the ransom to have the data recovered. Even companies and tech savvy people must use extensive resources to maintain backups for recovery or else they will lose valuable data, not mentioning average users. Unfortunately, not any recovery tool can effectively defend various types of ransomware. To address this challenge, we propose a novel ransomware defense mechanism that can be easily deployed in modern Windows system to recover the data and mitigate a ransomware attack. The uniqueness of our approach is to fight the virus like a virus. We leverage Alternative Data Streams which are sometimes used by malicious applications, to develop a data protection method that misleads the ransomware to attack only file 'shells' instead of the actual file content. We evaluated different file encrypting ransomware and demonstrate usability, efficiency and effectiveness of our approach.



\vspace{8pt}


  \end{abstract}

  \section{Introduction}\label{sec:introduction}

Ransomware is a type of malware  that has quickly become a rife means of attack due to the large monetary gains that are achievable. Specifically, ransomware operates by exploiting an attack vector such as a fake Adobe Acrobat update or a malicious email attachment to  deliver a payload to a victim's computer system. Once this payload is executed, the ransomware will limit the victim's access to the computer system by either locking the system or removing access to files through file encryption. Next, the victim will be instructed to pay a ransom. The ransom is typically in the form of crypto-currency and is required in order to regain access to their systems or files \cite{ransomware_study}.

\begin{figure}[!h]
    \centering
    \epsfig{file=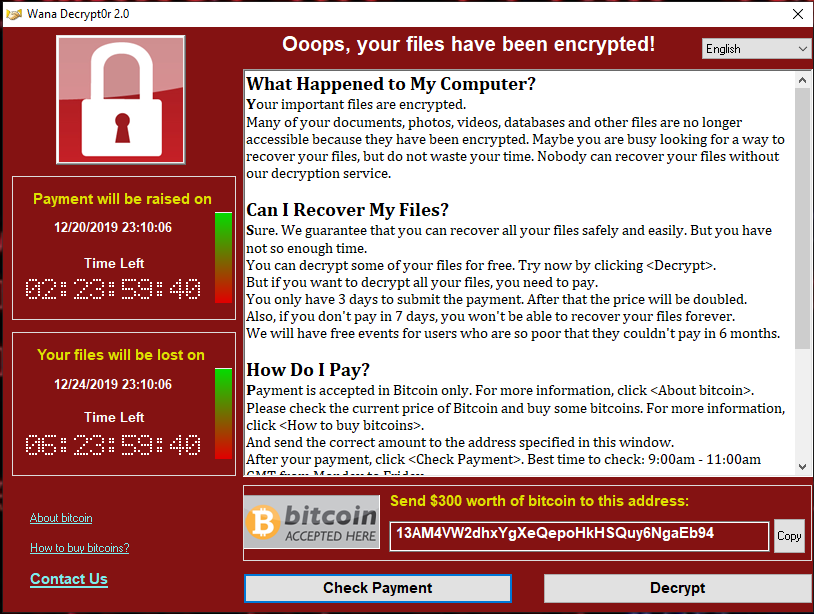,width=0.4\textwidth}
    \caption{WannaCry Ransomware}
    \label{fig:wannacry}
  \end{figure}

In this paper, we focus on tackling the file-encrypting ransomware that encrypts a list of files on the victim's computer. The popular modern adage, ``Data is gold", depicts exactly why ransomware is so effective. Regardless of industry, society has been steadily transitioning to a state of electronic and data dependence. This dependence along with several leading factors have created ideal conditions that have allowed ransomware to thrive. These key factors include the viability in financial gain and the difficulties of recovering from ransomware attacks. With a variety of crypto-currencies such as Bitcoin becoming more economically established, it provides an avenue for malicious actors to anonymously receive ransom payment without exposure. For the average computer user, it is nearly impossible for them to recover their data without payment. Even companies and tech savvy people must use extensive resources to maintain backups in hope of recovery or else they will lose valuable data. The difficulties of defending against ransomware and the ability of malicious parties accepting payments anonymously have led to disastrous results. Malicious parties from script kiddies to advance persistent threats (APT) have successfully leveraged ransomware campaigns against society. A prominent and possibly one of the most infamous ransomware is WannaCry.  WannaCry piggybacks off the NSA exploit, Eternal Blue, that was release by the APT Shadow Brokers. In the course of a day, WannaCry had infected an estimated 230,000+ computers and ultimately resulted in hundreds of millions to billions of dollars worth of damage. Figure \ref{fig:wannacry} is an example of WannaCry's ransom message. This message demands payment under threat of permanent file loss.

As aforementioned, the primary recovery strategy for a ransomware attack is to restore the system using backups \cite{flashguard, SSD_Insider, amoeba,sparse_space}. There are many different back up strategies, but they all require both storage space and time. Back up strategies typically deal with a trade off between data availability and time. Both businesses and home users must evaluate the costs and time associated with maintaining up to date and reliable back ups. Real time back ups are often far too expensive while other incremental strategies will still lead to overhead in storage and management. In the event of a ransomware infection, only the data from a back up will ensure data integrity. However, this too leads into the issue of identifying when the ransomware first infiltrated the system. If the ransomware remained dormant on the system prior to execution, then previous backups will have dubious integrity. Thus, even with a consistent backup routine, systems are still exposed and at risk to ransomware attack. Furthermore, some variants of ransomware \cite{ransomware_backups} have been documented to also encrypt backups. Ultimately, back ups are a last defense strategy for ransomware recovery.


Since existing recovery approaches are not always effective, efforts have been devoted to the development of ransomware detection techniques that aim to detect and stop the ransomware attack before it is able to encrypt all the files. These detection techniques resemble antivirus software. They try to identify potential ransomware attacks by identifying function calls in the code \cite{API_static_analysis, HAMPTON201844}, analyzing file access patterns and network traffic  \cite{locky_network_study}, or monitoring malicious changes to a small number of decoy files \cite{GOMEZHERNANDEZ2018389}. However, current approaches still encumbered with various limitations. Many of them reduce the computer system performance, require excessive storage space, or produce excessive false positives. Moreover, they are generally susceptible to novel forms of ransomware that have adapted to different detection algorithms and exhibit new behavior.


In order to  provide a more robust and effective defense against file-encrypting ransomware, in this paper, we propose a novel hybrid method called FREEDOM (Fast REcovEry and DetectiOn Mechanism),  which possesses both recovery and detection abilities. The uniqueness of FREEDOM is that it fights the virus like a virus. Specifically, we leverage the Alternative Data Streams (ADS) technique which is usually used by malicious applications, to  develop a data protection method that misleads the ransomware to attack only file `shells' instead of the actual file content. With our proposed system in place, users maintain ready access to their data without any perceived differences while having their data in a secure, defensive and recoverable posture. Additionally, our system is also resilient to
white-box attacks. FREEDOM is capable of detecting any illegitimate modification to the ADS style files by utilizing operations conducted in the operating system kernel that cannot be easily targeted by ransomware.  We have evaluated a number of ransomware samples and the experimental results demonstrate that our approach is an effective defense against ransomware targeting users' individual files. Our contributions are summarized as follows:
\begin{itemize}
\item We propose a novel defense mechanism called FREEDOM to fight against file-encrypting ransomware. Unlike many existing recovery mechanisms, FREEDOM introduces negligible performance overhead and requires nearly no additional space.


\item We not only consider current file-encrypting ransomware but also emerging ones. We propose a kernel-level monitoring algorithm which is capable of detecting potentially malicious changes to our proposed new file structure, thereby preventing ransomware targeting our method from causing data loss.


\item We have conducted extensive experiments on numerous real ransomware. The findings demonstrate both the efficiency and effectiveness of the proposed FREEDOM system.

\end{itemize}



The rest of the paper is organized as follows.   Section~\ref{sec:approach} presents our proposed FREEDOM system.  Section \ref{sec:exp} reports the experimental results. Section~\ref{sec:threat} analyzes the system security. Section~\ref{sec:related} reviews related work on ransomware defensive mechanisms. Finally. Section~\ref{sec:con} concludes the paper.

\section{The FREEDOM System}\label{sec:approach}

In this section, we introduce our proposed FREEDOM system that possesses several unique features. First, FREEDOM is capable of quickly recovering from existing file-encrypting ransomware attacks and guarantees zero file lost. Here, zero files lost refers to the ability to restore all the user's original files using the FREEDOM system. Second, FREEDOM introduces negligible operational overhead and  users will not notice the existence of FREEDOM. Third,  FREEDOM requires almost no any additional storage space. Finally,  FREEDOM is also robust against future ransomware that are aware of the mechanisms of FREEDOM.

The primary concept of the FREEDOM system is to obfuscate access to user files by utilizing Alternative Data Streams (ADS). This is similar to the process malware use to insert malicious code into ADS. However, unlike malware which only needs to execute its own code in ADS, our FREEDOM system faces the following new challenges:

\begin{itemize}

\item How can FREEDOM ensure that user experience remains consistent when interacting with the protected files? This is because ADS data is hidden from users and cannot be directly accessed using common applications like Acrobat or WORD. It is certainly not practical to require users to utilize special commands to access their files.

\item How can FREEDOM be resilient against attackers who know the internal mechanism of FREEDOM? In other words, if attackers know the users' files are hidden in ADS, can we prevent  attackers from directly encrypting the ADS or make the attack challenging enough such that it is effort prohibitive?

\end{itemize}

In order to address the listed challenges, the FREEDOM system consists of three major components:  an ADS file converter for file conversion, an ADS file restorer which restores user files after a ransomware attack, and a kernel-level driver monitor for detecting new types of ransomware. In what follows, we will elaborate on their technical details.

\subsection{ADS File Converter} \label{sec:linker}

\begin{figure}[!t]
    \centering
    \epsfig{file=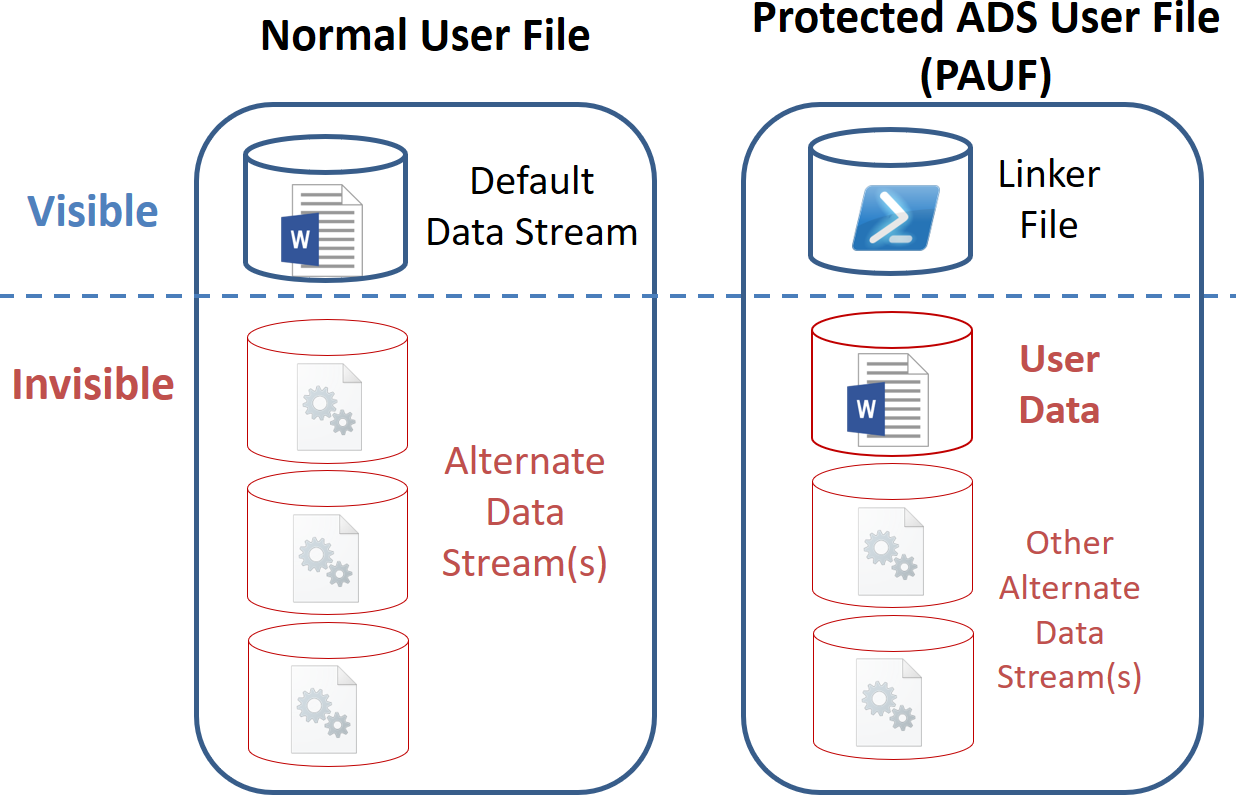,width=3.0in}
    \caption{Normal File vs. PAUF File}
    \label{fig:pauf}
    \end{figure}

Since Alternative Data Stream (ADS) is the building block of our system, we first briefly introduce its basic mechanisms. (ADS) is a file attribute in the NTFS file. The NTFS file system has a Master File Table (MFT) that stores file entries. Each entry in MFT describes a file. A file in NTFS is actually more than one file. It is essentially a container that has a collection of attributes stored as separate data streams (as illustrated in Figure \ref{fig:pauf}). These attributes include one default data attribute  and multiple named data attributes.  The default data attribute usually stores the actual file content (user data) or the index to the user data. This data attribute is the file that users are most familiar with and deal with. The other named data attributes are the so-called Alternative Data Streams (ADS), which are internal to the NTFS system and not visible to users without special tools. Not only invisible to users,  changes in ADS do not even affect the file size. Such obscure features make ADS an ideal vector for malware to insert their malicious code. While malware can hide the code in ADS, we can also take advantage of ADS' obscure features to hide user data such that ransomware will not be able to locate user data from traditional file searches. This is the starting point of the design of our FREEDOM system.

Although relying on the same ADS technique, the design of FREEDOM system faces many more challenges than malware. Malware only needs to hide one particular piece of malicious code in an ADS attached to a legal file and does not need to facilitate user access. On the contrary, our FREEDOM system not only needs to hide user data of various file types .pdf, .docx, .xlsx, and .jpg, but must also ensure that users can still operate these hidden files using the corresponding applications. In order to avoid encumbering the user with additional usage steps, all the file protection conducted by FREEDOM should be automatic and require no extra expertise. In summary, the FREEDOM system is required to be practical and user friendly. Ideally, when a user attempts to open their original file, the original file content should be displayed and editable through the associated application, e.g., a .docx file would be opened by Microsoft WORD, and a .pdf file would be opened by Adobe Acrobat.

To achieve the above goals,  we propose a new file structure as shown in Figure \ref{fig:pauf} along with an  ADS file converter that can automatically convert a user file to the new file structure. Specifically, the new file structure, called Protected ADS User File (PAUF), consists of two major components: a linker file and an ADS file with user data. The conversion process from a regular user file  to the PAUF is as follows. Given a normal file, say sample.txt, the content in sample.txt will first be copied to an ADS data stream named ``sample.txt:Data1.txt" using the following command.

{\tt cat sample.txt $>$ sample.txt:Data1.txt}

Next, we create a linker file  to provide access to the newly created ADS file. The linker file will replace the original user file 'sample.txt'. In order to allow users to perform normal file editions by double clicking the linker files which have now taken place of their original files, the linker files should contain commands that can automatically open the protected ADS user file. However, this is very challenging. First, we need to select an appropriate and widely available scripting language to support this automatic file opening task. For this, we adopt Windows PowerShell which is a task automation and configuration management framework from Microsoft, consisting of  a command-line shell and a scripting language. The second challenge is to design a succinct and generic algorithm to save storage space and support opening and editing different file types. As we know, user files have diverse types and each type of file needs to be opened with certain applications. It appears that a straightforward way is to use the PowerShell commands to identify the original user file extension, search its associated application, and then use that associated application to open the protected ADS user file. However, this does not work as expected because most applications are  not designed to interact with ADS files and  cannot save ADS files properly. For example, Microsoft WORD will report the error stating that  there are invalid characters in the file path. This is because ADS files contain the special symbol $:$ in the file path.  To overcome this obstacle, we propose a work-around approach which copies the ADS file to a temporary normal file to feed to the application and then saves the modified file back to the ADS file.

    \begin{figure}[!t]
    \centering
    \epsfig{file=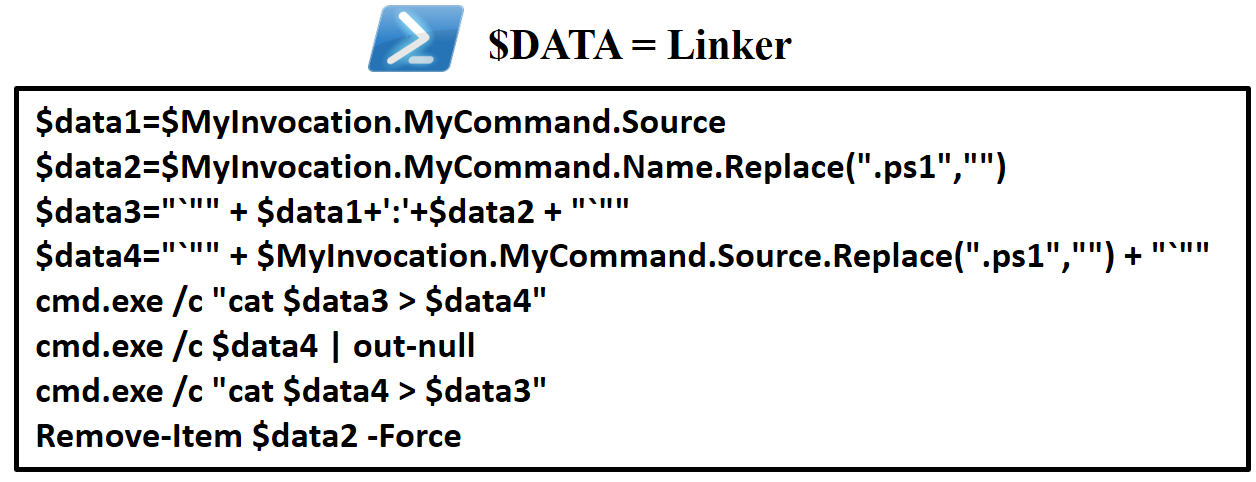,width=3.5in}
    \caption{Linker File Content}
    \label{fig:linker}
    \end{figure}

        \begin{figure}[!b]
    \centering
    \epsfig{file=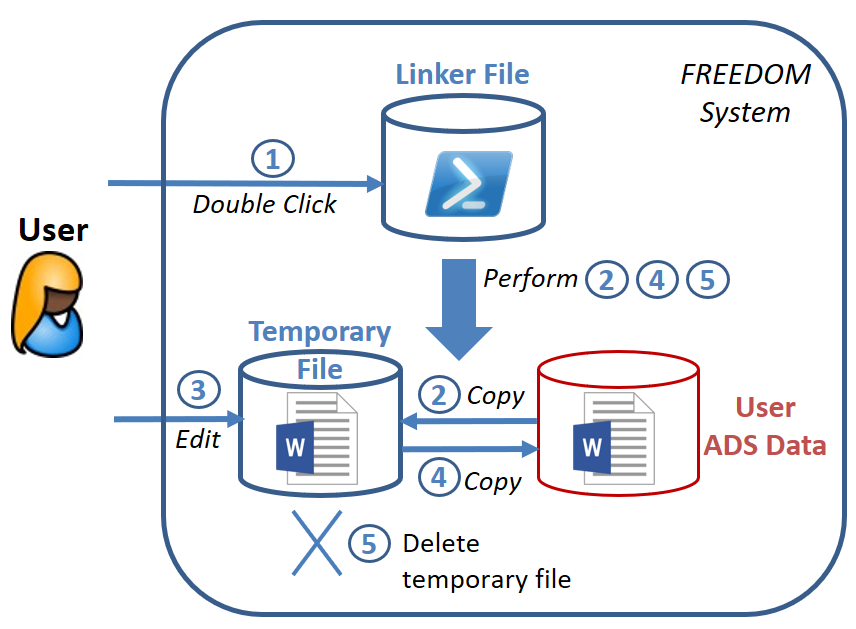,width=0.45\textwidth}
    \caption{Accessing the PAUF File}
    \label{fig:fileaccess}
    \end{figure}

Figure \ref{fig:linker} outlines the PowerShell commands in a linker file where \$data1, \$data2 and \$data3 refer to the linker file, the temporary file and the user's ADS file, respectively.  Figure \ref{fig:fileaccess} illustrates the data access procedure.  Specifically, upon being double clicked, the linker file will perform the following steps.  First, create a new temporary file with the same extension as the original user file. Next, copy the content of the protected ADS user file to the temporary file. Then, invoke the application program for the temporary file. The user can now see the content of the original file in the usual application and interact with the file normally.  After the user saves the content in the application and closes it, this will trigger our system to automatically  copy the content from the temporary file to  the protected ADS user file. Finally, the temporary file is deleted from the system.

It is worth noting that our linker file contains universal commands that work for any type of user files. This not only offers an easy deployment of our system but also  paves the foundation for detecting novel ransomware as elaborated in Section \ref{sec:detect}.


\subsection{ADS File Restorer}\label{subsec:storage}

In this section, we focus on how to recover from existing ransomware attacks. We will address the challenges of novel ransomware potentially designed to target our defense mechanism in the next subsection.

Due to existing file-encrypting ransomware only targeting normal files (i.e., the default data streams) and the fact that the normal files are now protected using our system, encrypting the linker file will not result in any user data loss. This is due the linker file obscuring access to the actual user content and thereby protecting the data. However, once a linker file has been encrypted by ransomware, users will be unable to conventionally access their data. Therefore, we still need a recovery mechanism to retrieve the user's original data following the attack.



To understand the file recovery process, we must be clear on how our proposed Protected ADS User File (PAUF) is stored. One of our design goals is to minimize the extra storage requirement. This allows our mechanism to be significantly more cost effective when compared to many existing solutions that require additional storage space for use data backups. In our FREEDOM system, we take advantage of the pre-allocated space within a NTFS entry to store the linker file so that the original user file size is unchanged.  Specifically, each file entry in the MFT is created with the same size, 1024 bytes, which are used to store file attributes including the default data stream and the alternative data stream (ADS). The default data stream is used to store user data while the ADS is used to store metadata such as the original file author. There are two ways of storing user data in the file entry. When the user data is small enough (much less than 1K bytes), the user data will be directly stored in the default data stream of the MFT file entry and be labeled as resident. In most cases the user data is larger than 1K bytes and the file entry only records a pointer to the disk location where the user data is stored. Such occurrences are labeled as non-resident. If a user file is non-resident, i.e., not stored in the file entry, there is always sufficient and unused space in the file entry to store our linker file due to the linker file only requiring 264 bytes. Specifically, the default data stream that is visible to users will be used to store the linker file. The ADS, that is not visible to users, is used to store the pointer to the user data.



If a user file is resident, the linker file is still stored in the default data stream while the user data can be handled in two different ways according to the data size. When the user file is very small and there is enough space to place both the linker file and user data in the file entry, the user data will be stored as an ADS in the file entry.  When there is not enough space to place both the user data and the linker file together in the file entry, additional space outside the file entry will be required to store the user data. In this case, the user data is less than 1K bytes, so the  additional space is less than 1K bytes as well. In reality, there are seldom user files less than 1k bytes in size. Therefore, the need of additional storage space in this occurrence can be considered negligible.

We now proceed to discuss how to restore user data after the system is attacked by a file-encrypting ransomware. Existing file-encrypting ransomware encrypts the default data streams in the file system. The actual user data is still safe in the ADS. To restore system back to pre-attack state, we simply carryout two steps. First, we locate all the ADS files that contain user data using the command ``dir /R".  Second, we replace each encrypted linker file with our original linker file. Following these steps, the users regain conventional access to their files. The recovered user data is free from any malicious effects of ransomware execution.

\subsection{Kernel-level Driver Monitor}\label{sec:detect}

Our file protection approach introduced in the previous sections is very effective against various versions of existing file-encrypting ransomware (as shown in our experiments as well). The design is based on the fact that the existing file-encrypting ransomware mainly target files that are visible through traditional directory searches. However, there exists the possibility that future ransomware may know of our implementation and attempt to target the two components of the PAUF file, the linker and the ADS file. These concerns are described as follows:


\begin{itemize}
\item  {\bf Attacking the linker file}: A new ransomware may attempt to modify the commands in the linker files to copy the user data in the ADS to their server and then delete or encrypt the users' ADS files.

\item  {\bf Attacking the ADS file}:  A new ransomware may search for all the ADS files and directly encrypt these ADS files which store the user data.

\end{itemize}

 \begin{figure}[!t]
    \centering
    \epsfig{file=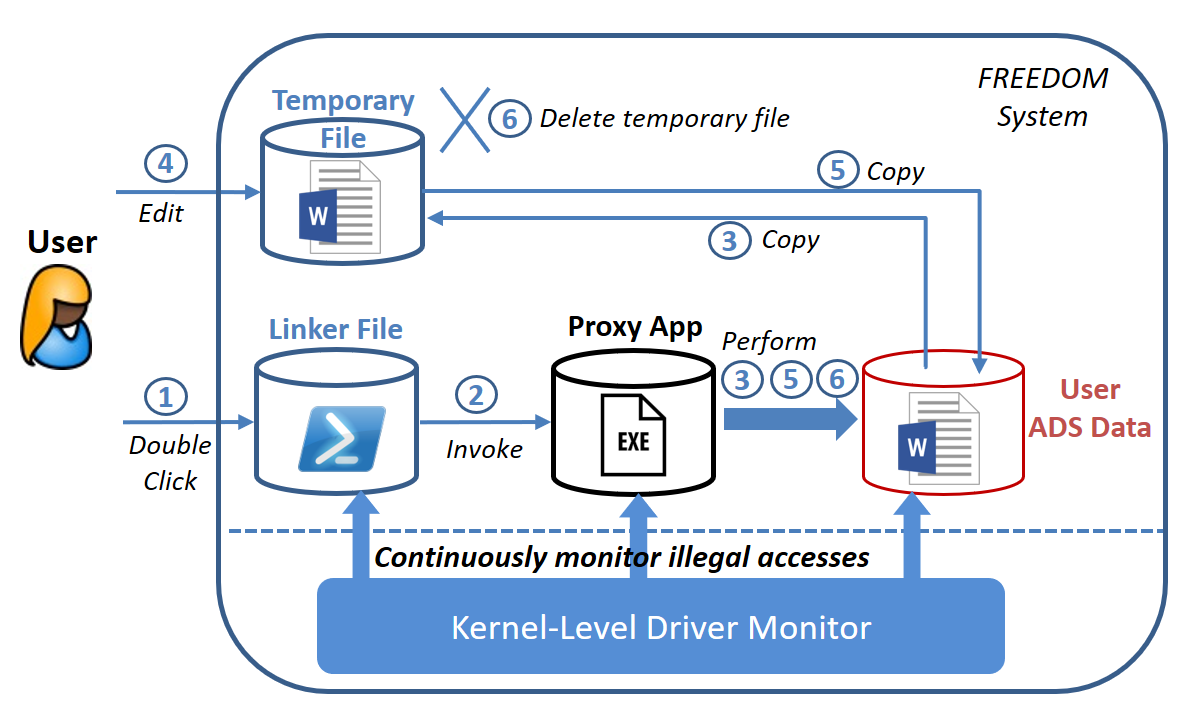,width=3.5in}
    \caption{FREEDOM System with Driver Monitor and Proxy Application }
    \label{fig:app_proxy}
    \end{figure}

In order to ensure resiliency against new ransomware, we  propose a kernel-level driver monitor to detect and prevent potential malicious modifications to the PAUF files. Specifically, we implement the  driver monitor using commands from Windows Driver Kit (WDK) to monitor the I/O operations on the linker files. Due to our linker files being uniform for all user data and that linker files are not supposed to be modified,  any write operations to the linker files will be deemed as malicious. Our driver monitor checks I/O operations to each process every two seconds and will terminate the process that attempts to write to the linker files. This way, we can quickly stop the attack to the linker files. We will also prompt an alert to the user detailing the potential threat. This strategy not only works for preventing future ransomware attacks, but also can be used to effectively terminate existing ransomware when they attempt to perform write operations, i.e., encrypting the linker files.

In order to ensure resiliency against new ransomware, we propose a kernel-level driver monitor to detect and prevent potential malicious modifications to the PAUF files. Specifically, we implement a driver monitor using Windows Driver Kit commands to view and store I/O operations. We define any files related to our PAUF files (linker and ADS), kernel-level driver, and FREEDOM system application as important. All files other than these will be discarded and not processed allowing our system to continuously monitor I/O operations while lowering our overhead. 

In the second attack scenario, our driver monitor will work in tandem with a more advanced version of the linker file. The advanced linker file will not contain the previous commands that copy the data between the temporary file and the ADS file. Instead, theses commands are now compiled as an executable file and the new linker file will call the executable file to perform the operations on the user data. The advantage of introducing the executable file as a proxy application is to enable the driver monitor to easily distinguish malicious accesses from legal accesses to the user's ADS files. In our system, legal accesses to user's ADS files should only be through the executable file invoked by linker files. Since the executable file is non-editable, it is impossible for the attacker to modify the executable file to perform encryption. In order to encrypt the user's ADS file, the attacker will have to use their own application to access the user file. However, any access from processes other than our designated proxy application will be deemed as malicious by our driver monitor.


Figure \ref{fig:app_proxy}  illustrates the PAUF file access procedure under the supervision of the driver monitor. I/O operations to the linker file, the proxy application, and the ADS files are continuously  monitored. All deletion and modification attempts to the linker files or the proxy applications are deemed malicious. Additionally, all deletion and modification attempts to the ADS file performed by applications other than the proxy are also deemed malicious. Not in the scope of already treated files, our system also monitors modification attempts to the file conversion methods and user interface program for our system. The  process that made the requested operation will be suspended. The  user will receive the alert to determine if they want to  terminate the process. This way, the user will also be informed of potential ransomware on their system.

\begin{figure}[!t]
    \centering
    \epsfig{file=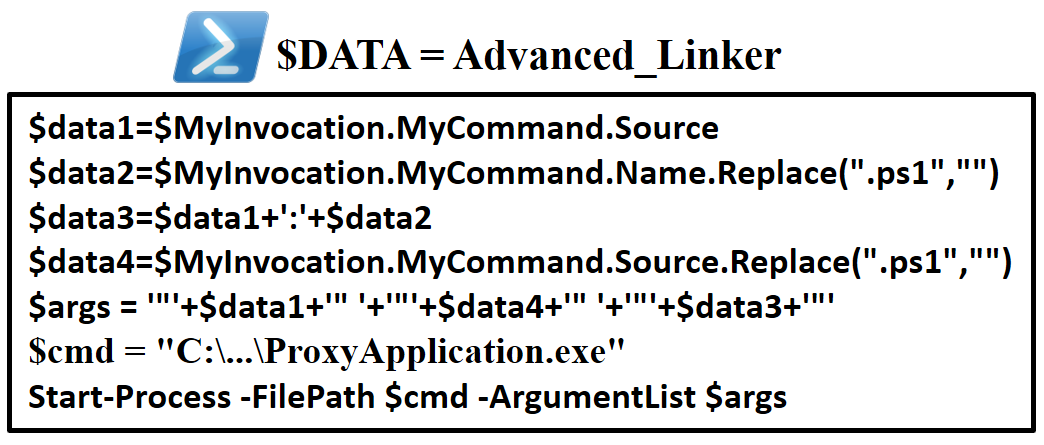,width=3.5in}
    \caption{Advanced Linker File Content}
    \label{fig:alinker}
    \end{figure}

Figure \ref{fig:alinker} shows the commands in the new linker file. The new linker file will first gather the full path of the linker file and the full path of the associated  ADS file. Then, the linker file will  call the proxy application, i.e., the  executable file, which handles further interaction with the user's ADS file. The proxy application will perform the same operations as the  linker file introduced in Section \ref{sec:linker}. Specifically, the executable file will copy the user data from the ADS file to a temporary file with the same extension as the user's original file so that the temporary file can be opened and edited by its corresponding normal application. This means that means normal applications will not directly access the user's ADS files. By allowing only our proxy application and other trusted Windows essential applications to modify the ADS files, our driver monitor can easily identify malicious applications that attempting to access the user data.

It is worth noting that our driver monitor is substantially more accurate and effective than other  driver-based monitoring techniques \cite{NAGuard, cryptolock, shieldfs, unveil, 2endFox, NN_defense}. Although existing driver-based monitoring approaches also look at file changes and monitor operations such as file writes, file reads, file deletions, file creations, etc., they can only  make educated guesses when determining if a process is malicious through analysis of the system's historical normal behavior. They cannot guarantee to always detect a malicious process. Moreover, for their drivers to identify malicious processes, malicious behavior must be shown which allows  ransomware to execute unhindered until it passes a threshold that identifies it as a malicious process. This causes a number of files to be encrypted or lost before their drivers can stop the ransomware. Our FREEDOM system avoids these problems by design. We have a unified linker file and a generic proxy application for all the user data. This design consistency ensures high accuracy in classifying normal and malicious operations on the user data  as described in our detection algorithm. In summary, any illegal operations are guaranteed to be detected by our driver monitor.

\begin{figure}[!t]
    \centering
    \epsfig{file=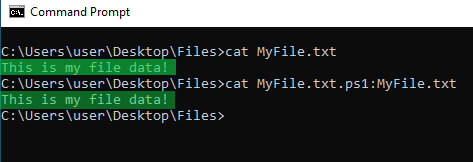,width=0.45\textwidth}
    \caption{Normal File and PAUF File Before Ransomware Attack}
    \label{fig:startfilesdata}
  \end{figure}

\begin{figure}[!t]
    \centering
    \epsfig{file=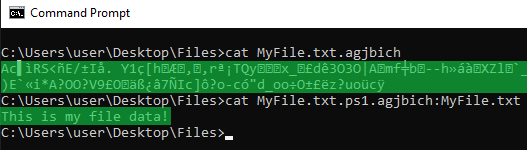,width=0.45\textwidth}
    \caption{Normal File and ADS File After Ransomware Attack}
    \label{fig:adsstillthere}
  \end{figure}

 Figures \ref{fig:startfilesdata} and \ref{fig:adsstillthere} show a simple example of how our FREEDOM system reacts to a ransomware called GrandCrab. GrandCrab is one of few existing ransomware that encrypt PowerShell files - the same file type as our linker files. In this example, we first create a file called  $MyFile.txt$ which contains a simple sentence ``This is my file data!". Next, we  convert the file into our PAUF format.  Figure \ref{fig:startfilesdata} shows that the ADS file contains the same data as that in the original user file. Finally, we allow the GrandCrab ransomware to run.  Figure \ref{fig:adsstillthere} shows the file content (highlighted) after the ransomware attack. Observe that normal files, such as the original $MyFile.txt$, have now been encrypted by the ransomware. However, our converted ADS file remains uninfected. Moreover, if the driver monitor  is turned on, it will detect the ransomware's attempt to encrypt the linker file and alert the user so that the user can decide to terminate the malicious program.

\section{Performance Study}\label{sec:exp}


\subsection{Usability Evaluation}

\begin{figure}[!t]
    \centering
    \epsfig{file=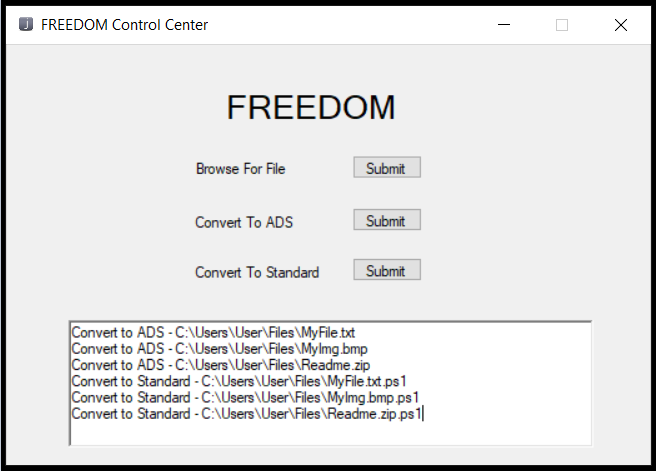,width=2.8in}
    \caption{FREEDOM System GUI}
    \label{fig:GUI}
 \end{figure}

We have implemented all the components in the FREEDOM system including file conversion, file restoration, the proxy application, and the kernel-level driver. With our system modifying the type and storage of all user files, ease of use must be considered. For this, we developed a GUI program (as shown in Figure \ref{fig:GUI} that provides the functions to convert any  file to our proposed PAUF file format, revert PAUF files back to the normal format, and manage the driver for the system. It is worth noting that this does not mean that a user needs to perform all the functions each time he/she needs to access the file. In practice, the user only needs to perform the file conversion function once to convert files he/she would like to protect. Once the file conversion is completed, the user can double click the converted files and edit them  just like normal files as many times as needed. Therefore, our system is much more convenient for users than many security software that requires the user to  first decrypt or type in the password and later encrypt the file each time the user edits the file.

The only time that the file conversion needs to be conducted again is when the user wants to switch the application that is used to open a file. In our current system, a linker file associates each file type with one frequently used application for that file type. If a user wants to change the application for a particular file, the  linker file needs to be recreated to achieve that. This is  similar to the operation in Windows  whereby a user needs to right-click a file to set a new application to associate with a file type.

To sum up, although our system does not support the other way of file opening by  opening the application first and then selecting the file, our system does not affect the user's file editing operations inside the application, nor affecting the user's choice of applications. Additionally our system can be implemented on a case by case method, files that do not need defense or do not work easily with our system do not need to be converted and defended by our system. With this in mind our system is best suited to defend files that easily work within our system. Such use cases is defending Microsoft Office Suite files on employee computers where they may not be fully trained to deal with cyber security issues or may unknowingly infect their computer. In this case important documents are protected between mandatory backups saving important documents on an employee's computer.


\subsection{Effectiveness Evaluation}

To evaluate the effectiveness of our FREEDOM system, we test it against 8 common file-encrypting ransomware: WannaCry, GandCrab, CryptoLocker-v3, TeslaCrypt, Vipassana, Locky, JigSaw, and Cerber. They were downloaded from the open source malware repositories \cite{thezoo, virusshare}. We would like to mention that it is possible that the versions collected by these repositories are variants of the original ransomware, but that should not affect the evaluation of the effectiveness of our system.

In the experiments, we installed our FREEDOM system in a virtual machine with a recently patched Windows 10 operating system. This environment emulates a typical user system. The test environment contains traditional files of various formats including: .txt, .jpg, .pdf, .docx, .xlxs, .pptx. We first run the FREEDOM system to convert the traditional files to our PAUF file format. Then, we disable the native antivirus and other defensive measures to evaluate the sole effect of the ransomware infection under the defense of our FREEDOM system.

In each round of experimentation, we have a freshly installed system as described above. In each round we run a single ransomware and then examine the following items:
 \begin{itemize} \itemsep=0pt

\item Has any user's ADS file been encrypted?

 \item Has any linker file been encrypted when the driver monitor is turned off?

 \item Has any linker file been encrypted when the driver monitor is turned on?

 \item Has the ransomware been terminated by our driver monitor?

 \item Are the user data recoverable after the attack?   Here, the ability to recover is defined  as being able to restore the users' access to their data to the normal state without using software other than our FREEDOM system. No drive recovery techniques or intensive inspection will be used for this process.

 \end{itemize}

\begin{figure}[!ht]
    \centering
    \epsfig{file=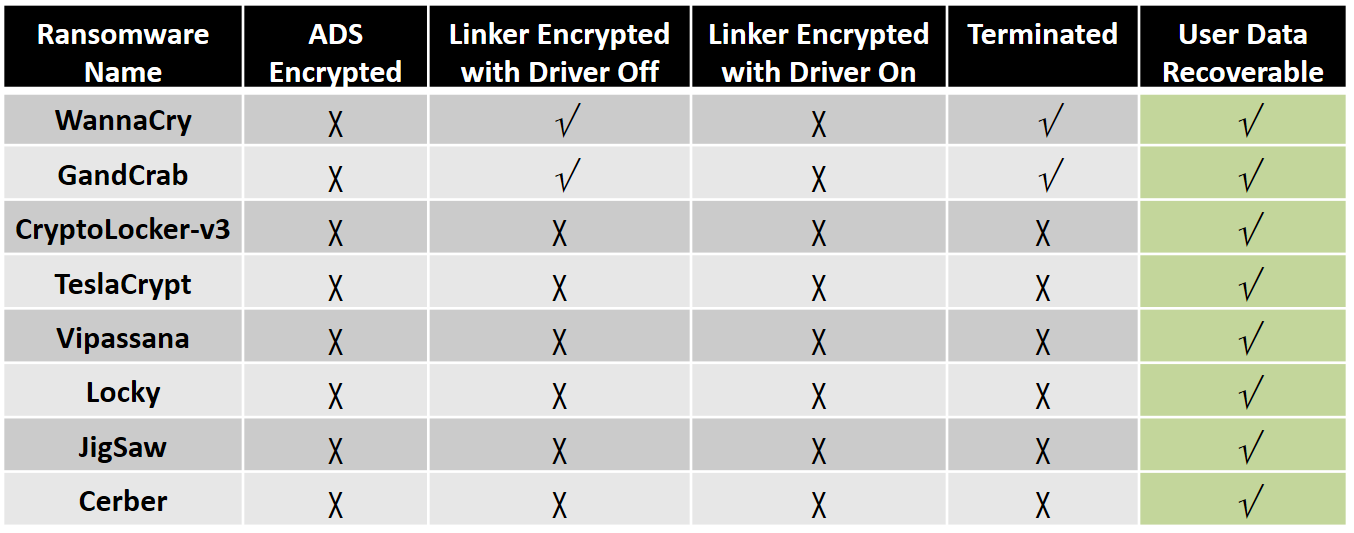,width=3.5in}
    \caption{Ransomware Tests}
    \label{exp:attack}
  \end{figure}

Figure \ref{exp:attack} shows the experimental results. As expected, none of the ransomware being tested encrypted the user's ADS data. This is because a majority of ransomware have a list of file extensions to encrypt and these file extensions are usually those common file types that users often use to store their personal data. From gathered research, many of the existing file-encrypting ransomware do not intend to restrict the victim's usage of the computer system. Therefore, they do not target ADS files which are mainly for system usage.

We do see that some ransomware such as WannaCry and GrandCrab target a larger list of file extensions that also include PowerShell files, e.g., our linker files. In the experiments, when we use the original version of the linker file (Section 2.1) without enabling the driver monitor, WannaCry and GranCrab will encrypt these linker files. Once we switch to the advanced linker file version along with the driver monitor, none of the linker files are encrypted by the ransomware. Additionally,  any ransomware that attempts to encrypt linker files (e.g., WannaCry and GranCrab)  have also been terminated by our driver monitor. This proves the effectiveness of our driver monitor in detecting the malicious operations on linker files.  Even though we do not have any future ransomware to test, the observed ability  of the driver monitor in terminating ransomware that attempt to encrypt the linker file indicates that our driver monitor will be able to monitor malicious operations on the proxy application and ADS files in a similar way. This demonstrates  the potential of our system in defending against future ransomware. It is worth noting that the success is attributed to not only the driver monitor, but also the new PAUF file format which enables clear classification of malicious behaviors.

It is not surprising to see that the driver monitor did not terminate ransomware that does not encrypt linker files (or ADS files). This is  because they did not trigger the monitoring process which looks for malicious operations on the linker files, the proxy application, and the ADS files. In that case, it is inconsequential whether such ransomware is terminated since none of the user's critical files (i.e., PAUF files) has been encrypted by the ransomware and no recovery is needed at all. In the event the linker files are encrypted or damaged, the recovery process in our FREEDOM system is simple. Our system can quickly replace the encrypted linker files (if any) with original linker files to assist users in regaining access to their data.

\subsection{Efficiency Evaluation}

Besides effectiveness, we also evaluate the possible overhead incurred by our FREEDOM system during the file conversion and normal file operations. We create files with different file extensions and sizes ranging from 10KB to 1GB. Note that our approach works for any type of files theoretically. In the experiments, file tested are the common types including .txt, .jpg, .docx, .pptx, .xlxs, .py, .cpp, .mp4, .mp3.

\begin{figure}[!ht]
    \centering
    \epsfig{file=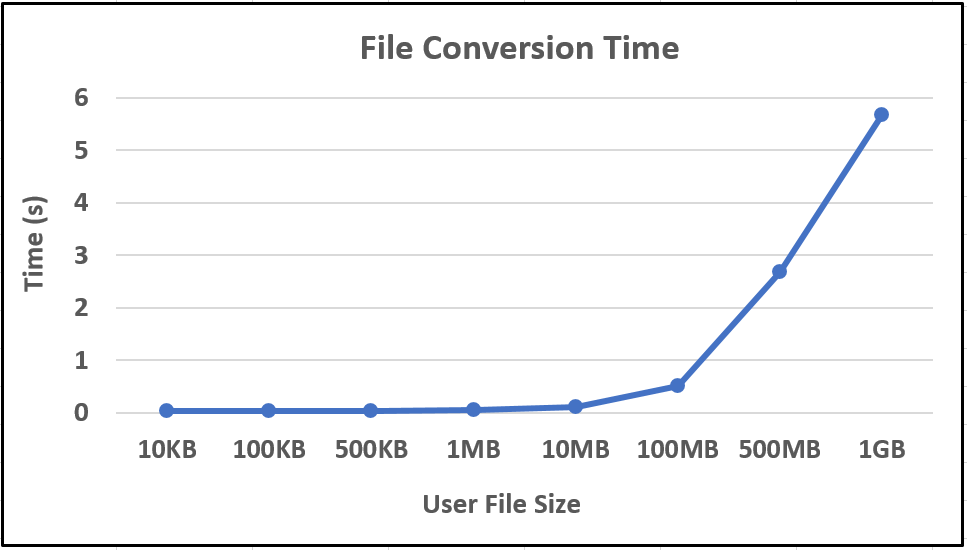,width=2.8in}
    \caption{Average Time Taken to Convert a Normal File to a PAUF File}
    \label{exp:convert}
 \end{figure}

Figure \ref{exp:convert} shows the average time for converting a single file of different types on the virtual machine (8-core 4.2GHz CPU and 4GB memory) to our PAUF file format. Observe that the file conversion is very fast. It takes less than half  second to convert files smaller than 100MB. When it comes to larger files such as videos, the conversion still can be done in a matter of seconds (e.g., around 6s for 1 GB file). The conversion time is linear to the file size mainly due to the need to copy the original file content to the ADS. The creation of the small linker file is almost negligible. Since the majority of files used daily, such as word documents and pictures, are typically less than 100MB, we expect the conversion during daily use will incur little overhead.  Additionally,  the conversion can be done in the background and will not inhibit user computer usage.

\begin{figure}[!t]
    \centering
    \epsfig{file=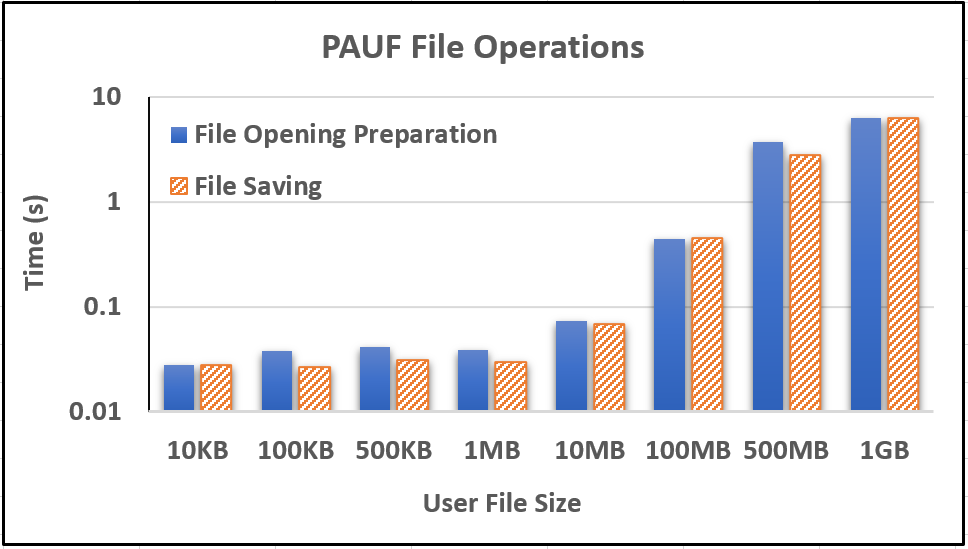,width=2.8in}
    \caption{Extra Time of Opening or Saving a PAUF File}
    \label{fig:file_opening}
 \end{figure}

Next, we check if our system will affect user experience with respect to opening and saving a file. Figure  \ref{fig:file_opening} shows the extra time needed to open and save a PAUF file compared to the time taken to open and save a normal file of the same size. As shown in the figure, preparing the opening and saving a PAUF file require no more than half second of additional time when the file size is less than 100MB. When the file size increases, the needed preparation and saving time also increase, but they are still very fast (around 6s for a 1GB file). This demonstrates the feasibility of our approach in practice.  The extra time is mainly introduced by the creation of the temporary normal file during the PAUF file opening and saving processing.  Recall that the FREEDOM system uses the temporary files to automate the user's access to the protected ADS data. The temporary files are created based on the user's ADS file content with the goal to offer the users consistent interaction similar to how they interact with normal files. Moreover,  the file saving proceeds in the background and users do not even have to experience this delay.

\section{Security Analysis}\label{sec:threat}


There are three common types of ransomware: (i) screen locker which locks the victim's computer screen; (ii) hard drive encrypter which encrypts entire hard drives in the victim's computer; (iii) file encrypter which encrypts a list of the victim's personal files. Our proposed FREEDOM system is designed to mitigate the existing file encrypters and future file encrypters.

While under the attack from existing file-encrypting ransomware, our FREEDOM system will attempt to detect and terminate the attacks via the kernel-level driver monitor as early as possible. As shown in the experiments, we successfully terminated  7 out of 8 ransomware. Regardless of whether the ransomware was able to be terminated, our file recovery mechanism will guarantee consistent and successful results. Since existing file-encrypting ransomware do not encrypt ADS files, all the user data is protected by our FREEDOM system. We only need to replace the encrypted linker files with the uniform linker file to restore users' access to their data. In summary, our FREEDOM system guarantees no file loss from existing file-encrypting ransomware.

As previously discussed, a challenging problem will be to successfully defend against ransomware developed to target the new file structure in our Freedom system. We classify the potential attacks into the following categories:
\begin{itemize}

\item {\bf Attempt to encrypt the user's ADS file}: With the FREEDOM system in place, the user's data is hidden in the ADS files. Therefore, a new ransomware may try to find all the ADS files and encrypt them. However, such attack will not be successful since user's ADS files are only allowed to be accessed by our proxy application. Any process other than our proxy application that attempts to modify the ADS files will be suspended by our kernel-level driver monitor. An prompt will be shown to the user to decide whether this process should be terminated.

\item {\bf Attempt to modify or replace the proxy application}: Since directly targeting the user data in the ADS file is no longer a viable method of attack, attackers may pivot to targeting the proxy application instead. However, since the proxy application is an executable file, which is non-editable, the attacker will not be able to  modify the proxy application's code. Moreover,  any modification to the proxy application can be easily blocked by our driver monitor since the proxy application is not expected to be modified after creation.  Alternatively, if the attacker attempts to fully replace the original proxy application with a malicious one, the replacement attempt will also be noticed by our driver monitor since the proxy application is not supposed to be deleted after creation. Any potential malicious replacement or modification targeting the proxy application will be detected and prevented by our driver monitor.

\item {\bf Attempt to modify or replace the linker files}: Another possible attack is to target the linker files.  With the FREEDOM system in place, user's data is hidden in the ADS files and is only allowed to be modified through double clicking the corresponding linker file. For new ransomware attempting to encrypt the user's ADS files, they may try to modify the original commands in the linker files from normal execution to file encryption. However, such attack will not be successful because any modifications to the existing linker  files will be detected by our kernel-level driver monitor since the linker files are not supposed to be modified or deleted after they are created. This also means attackers cannot delete the original linker files and replace them with malicious ones.


\item {\bf Attempt to uninstall the kernel-level driver monitor}: The last effort new ransomware may attempt is to uninstall our kernel-level driver monitor so that they can encrypt user's ADS data without being detected. This will require the ransomware to perform some type of privilege escalation and gain the administrator rights.  Once the ransomware obtains administrator rights, it can effectively perform any operations it wants. In such an event, the ransomware will be able to operate unhindered and other ransomware prevention and mitigation strategies will be ineffective due to the possibility of being uninstalled. Fortunately, 90\% of existing ransomware strains do not seek administrator rights \cite{ransomware_admin_fact}. We argue that although our FREEDOM system cannot prevent ransomware that have the capability to gain administrator rights, the requirement of administrator rights will significantly increase the development cost of ransomware capable of targeting the FREEDOM system.

\item {\bf Attempt to encrypt the files during normal usage}: This method of attack has not been seen in any of the tested ransomware samples. This attack method would involve ransomware gaining remote access to the user's computer and changing the files through our system by interacting with normal programs like the standard user would. As our system treats this type of access as normal user access, this would be seen as allowable. If an attacker went through each file on the computer and changed them this way, they could encrypt the entire file system. This method is extremely slow, is easily seen by any user on the computer, and is not the normal mode of operations associated with ransomware. Generally this attack would only be done in a targeted attack in which case ransomware is most likely not the best usage. This type of attack is considered outside the scope of our defense scheme as we are defending against general automated forms of ransomware.

\end{itemize}

To sum up, our FREEDOM system is  robust against existing file-encrypting ransomware and future ransomware that do not seek administrator rights on the victim's computer.

\vspace{4pt}
\section{Related Work}\label{sec:related}

Aside from user awareness training, the technological way of defending file-encrypting ransomware  can be classified into three main categories: (i) backups; (ii) code analysis; and (iii) dynamic monitoring.

Data backup is a commonly adopted strategy to recover user data from malware and ransomware attacks \cite{backups}.  Several existing ransomware defense methods \cite{flashguard, SSD_Insider, amoeba,sparse_space} utilize different strategies to create secure backups on the system and prevent the backups from being encrypted by ransomware.  While creating data backups are an effective solution to ransomware attacks, it comes with heavy storage costs since a duplicate of user data needs to be maintained. Additionally, if backups are not performed on regular basis, users will still lose valuable data after the ransomware attack. Unlike any of the existing backup methods, our proposed FREEDOM system neither needs to store duplicates of user data nor require users to perform any regular backup.

Another strategy to defend the system against ransomware attack is code analysis \cite{static_and_dynamic_anlysis_exporation, API_static_analysis, dynamic_analysis_suvery}. Code analysis relies upon static and dynamic analysis in the effort to draw conclusions from an artifact based on the native characteristics of ransomware. This is similar to an anti-virus software that checks the hash of an executable against a known database of malware and discerns if this may be a malware. These characteristics can also be used to train classifiers or set up entropy measurements.  However, the detection accuracy of static analysis approaches are not guaranteed. This is evident when examining anti-virus' poor detection rate regarding novel or malware variants.



A more advanced defense mechanism  is dynamic monitoring \cite{ransomware_survey}. Since ransomware typically needs to  access user files and modify them in a small time frame, they make incredible amounts of system noise that is uncommon for any normal use.  Based on these observations, various dynamic analysis algorithms \cite{cryptolock, honeypot_detection, 2endFox, NAGuard, thread_traversal_detection}  have been proposed to detect potentially malicious behavior in the system. Upon detection of abnormal activities, the suspected process will be terminated to mitigate the impact on the system.  For example, NAGuard \cite{NAGuard} operates by monitoring system I/O operations, analyzing the data, and terminating the suspicious processes. Besides monitoring the system behavior,  Taylor et al \cite{sensor_detection} propose to monitor hardware sensor behavior.  They studied the relationship between increasing and decreasing sensor readings to draw conclusions on what type of task the system is currently preforming. Another thread of works on dynamic analysis monitors network-based traffic. As ransomware usually utilizes the Internet to get keys for encryption, Cabaj et al. \cite{networkcryptowall} and Ahmadian  et al. \cite{networkdetection} leverage network monitoring to identify ransomware while it makes calls. However, existing ransomware detection mechanisms have two common limitations. First, their detection accuracy is not guaranteed because the classification of malicious behavior is based on some heuristic functions. There may be either false positives or false negatives. Also, for the classifier to work, it  takes time to collect and learn user's normal behavior. Second, for the detectors to identify malicious processes, malicious behavior must be shown. This allows ransomware to execute unhindered until it passes a threshold that identifies it as a malicious process. This causes a number of files to be encrypted or lost before the detector can  stop the ransomware. These detection methods cannot terminate suspected  behavior too early since terminating processes incorrectly could cause problems to the user or lead to system instability. In order to avoid sacrificing user files for the detection,   Continella et al. \cite{shieldfs}  propose to  couple the procedures of detection and recovery. Kharaz et al. \cite{unveil} use a similar method, but create an artificial environment for the detection.  Even though such hybrid approach can help protect user data during the detection, they still may not be able to detect the new ransomware which does not carry any known features learned by the detection system.


Compared to the previous works, our FREEDOM system does not have any of the limitations metioned above. Attributed to our new file structure (i.e., the linker file and the ADS user file),  we guarantee 100\% detection accuracy in terms of identifying ransomware that  attempt to modify users' protected ADS files.  We  do not sacrifice any user protected data in the detection process. Our system is resilient against both existing and future file-encrypting ransomware that do not gain administrator rights.


\vspace{4pt}

  \section{Conclusion}\label{sec:con}

  In this paper, we propose a novel approach called FREEDOM to fight against  file-encrypting ransomware. The biggest advantages of FREEDOM are that it does not require extra storage space to save user data and introduces negligible overhead. The design of FREEDOM leverages the stealth ability of Alternative Data Stream (ADS) to form a new structure of the user data which will not be infected by existing file-encrypting ransomware. Along with a novel kernel-level driver monitor, our FREEDOM system also has the future potential to defeat new file-encrypting ransomware that are aware of our underlying mechanisms. This driver monitor is also unique in that it ensures 100\% detection accuracy which is attributed to the new data structure in the FREEDOM system. We show that our system can successfully safeguard user data against real ransomware and our experimental results also demonstrate the efficiency of our system.


\bibliographystyle{IEEEtran}
  \bibliography{ref}

  \end{document}